\documentclass[11pt]{article}

\usepackage[margin=1in]{geometry}
\usepackage{lscape}
\usepackage{graphicx}
\usepackage{amsfonts}
\usepackage{amsthm}
\usepackage{amsmath}
\usepackage{amssymb}
\usepackage{dsfont}
\usepackage{bm} 
\usepackage{enumerate}
\usepackage{mathtools}
\usepackage{adjustbox}
\usepackage{comment}
\usepackage{tabularx}
\usepackage{threeparttable}
\usepackage{authblk}

\usepackage{soul}

\usepackage{natbib}
\usepackage{multirow}
\usepackage{booktabs}
\usepackage{color}
\usepackage{float}
\usepackage{ragged2e}
\usepackage{setspace}
\usepackage{caption}
\usepackage{subcaption}
\usepackage[toc,page]{appendix}

\makeatletter
\renewcommand*\env@matrix[1][*\c@MaxMatrixCols c]{%
  \hskip -\arraycolsep
  \let\@ifnextchar\new@ifnextchar
  \array{#1}}
  \makeatother

\newcommand{\tcr}{\textcolor{red}}

\newcommand{\E}{\mathbb{E}}

\newcommand{\U}{\mathbf{U}}

\newcommand{\Z}{\mathbf{Z}}

\newcommand{\z}{\mathbf{z}}

\newtheorem{theorem}{Theorem}
\newtheorem{lemma}{Lemma}

\newtheorem*{defn*}{Definition}

\newtheorem{example}{Example}
\newtheorem{remark}{Remark}

\newcommand\independent{\protect\mathpalette{\protect\independenT}{\perp}}
\def\independenT#1#2{\mathrel{\rlap{$#1#2$}\mkern2mu{#1#2}}}

\definecolor{candypink}{rgb}{0.89, 0.44, 0.48}
\definecolor{hgreen}{rgb}{0.21, 0.37, 0.23}

\usepackage{hyperref}
\hypersetup{
    colorlinks = true,
    citecolor = {blue},
}

\usepackage{natbib}
\begin{document}

\title{Identifying Key Influencers using an Egocentric Network-based Randomized Design}

\author[1,2]{Zhibing He} 
\author[3]{Junhan Fan} 
\author[4]{Ashley Buchanan}
\author[1,2]{Donna Spiegelman}
\author[1,2]{Laura Forastiere \thanks{Corresponding author. Email: laura.forastiere@yale.edu}} 
\affil[1]{Department of Biostatistics, Yale University, New Haven, U.S.A.}
\affil[2]{Center for Methods in Implementation and Prevention Science, Yale University, New Haven, U.S.A.}
\affil[3]{Roche Canada, Mississauga, Canada}
\affil[4]{Department of Pharmacy Practice and Clinical Research, University of Rhode Island}

\date{\today}

\maketitle

\begin{abstract}
Behavioral health interventions, such as trainings or incentives,  are implemented in settings where individuals are interconnected, and the intervention assigned to some individuals may also affect others within their network.
Evaluating such interventions requires assessing both the effect of the intervention on those who receive it and the spillover effect on those connected to the treated individuals.
With behavioral interventions, spillover effects can be heterogeneous in that certain individuals, due to their social connectedness and individual characteristics, are more likely to respond to the intervention and influence their peers' behaviors. 
Targeting these individuals can enhance the effectiveness of interventions in the population.
In this paper, 
we focus on an Egocentric Network-based Randomized Trial (ENRT) design, wherein a set of index participants is recruited from the population and randomly assigned to the treatment group,
while concurrently collecting outcome data on their nominated network members, who remina untreated.
In such design, 
spillover effects on network members may vary depending on the characteristics of the index participant.
Here, 
we develop a testing method, the Multiple Comparison with Best (MCB), to identify 
subgroups of index participants whose treatment exhibits the largest spillover effect on their network members.
Power and sample size calculations are then provided to design ENRTs that can detect 
key influencers.
The proposed methods are demonstrated in a study on network-based peer HIV prevention education program, providing insights into strategies for selecting peer educators in peer education interventions.

\end{abstract}

\noindent%

\section{Introduction}\label{sec:intro}

Causal inference has become a pivotal aspect of research in the health and social science, allowing researchers to establish causality between interconnected variables and guide decision-making. The Stable Unit Treatment Value Assumption (SUTVA), introduced by \citet{Rubin1974, rubin2005causal}, is a fundamental concept in causal inference. SUTVA assumes that the treatment or intervention assigned to one unit does not impact the outcomes of other units \citep{Rosenbaum2007}, and serves as a foundational principle in most of the conventional causal analysis approaches. However, 
this assumption is often violated when units are interconnected with other units through social or physical interactions. 
For example, in education, students participating in tutoring programs may exert an influence on the academic performance of their classmates through information sharing and peer interactions \citep{ murnane2010methods}.
Similarly, behavioral interventions such as training sessions designed to reduce health risk behavior (e.g.,  unprotected sex, alcohol and drug use, smoking), may have an effect on individuals beyond those receiving the intervention \citep{Buchanan2018}. In particular, individuals who change their behaviors in response to the received training are likely to influence their social connections to do the same.

To leverage and amplify these mechanisms of peer influence, behavioral interventions often rely on peer-based strategies, where specific individuals are trained to prevent risky behaviors and encouraged to disseminate knowledge and behavioral change among their social networks. 
Such peer education interventions have been effective in increasing HIV knowledge and reducing risk behaviors among both the peer educators receiving the training and their network members 
\citep{cai2008long,aroke2022evaluating}. Causal inference methods have been developed for settings with spillover among units have been also used to evaluate peer education interventions, disentangling the  direct effect of the training on treated individuals and the spillover effect on their social connections \citep{Buchanan2018, fang2023design, chao2023estimation}.

Much of the existing research on causal inference with spillover, also known as interference, focuses on the partial interference assumption, which allows for interference within groups, such as villages and schools, but not across them \citep{HudgensHalloran2008, TchetgenVanderWeele2012}. However, recent efforts have aimed to relax this assumption, allowing for more intricate interference occurring within networks of interconnected units 
\citep{SofryginLaan2016,Forastiere2020,yuan2021causal}.
Under network interference, the potential outcomes of one unit are influenced not only by its individual treatment but also by the treatments received by other units directly or indirectly connected within the network. 
For example, interventions or preventive measures for infectious diseases are likely to spillover across contact networks by reducing the risk of infection and transmission. Similarly, in behavioral interventions designed to promote healthy behaviors, individuals who change their behaviors in response to an intervention are likely to influence their social connections to do the same \citep{Buchanan2018}.

Peer influence may be highly heterogeneous, with certain individuals  being 
more influential than others. Often, individuals with more social connections are selected as peer educators \citep{grunspan2014understanding}.
 However, individuals who are more central in a network may vary inthe strength of their influence depending on their unique traits \citep{Kelly2006}. Identifying those who can effectively drive behavioral change among their social connections is crucial. This knowledge allows for tailored interventions, 
 delivering the training and selecting peer educators
 not only based on network position but also on individual's likelihood to adopt the desired behavior and, in turn, influence their network peers. 
\citet{lee2023finding} 
proposed a definition of causal influence as the effect achieved on the average outcome of the overall network if each specific group was treated, but do not provide any methods for identifing the most influential groups. On the other hand, 
\citet{qu2021efficient}
proposed a generalized augmented inverse propensity weighted estimator to assess the heterogeneity of direct and spillover effects under a partial interference setting.
Another contribution in this area is that of \citet{bargagli2020heterogeneous}, who developed a tree-based algorithm to assess the heterogeneity of intervention and spillover effects, with the latter defined in terms of susceptibility to the treatment of others as opposed to the strength of influence on others.

In this paper, we propose to use an Egocentric Network-based Randomized Trial (ENRT) design, used to evaluate peer education interventions. In this design, a set of index participants whose social networks are sufficiently separated  is randomized to either the control group or  the intervention group. In the latter, index participants receive training sessions on risk behaviors and are given the role of peer educators for their network members. In both the control and treatment group, index participants are asked to nominate a set of social network members (e.g., sex partners, drug partners), whose behavioral outcomes are collected in a follow-up survey. Network members do not directly receive the treatment, i.e., the training sessions, but may be exposed to the information received by their index participant indirectly. 
This setup allows for a straightforward estimation of the average spillover effect of the treatment of index participants on their network members' outcomes, as well as the exploration of its heterogeneity.
Analyzing the variation of these spillover effects in relation to index-level characteristics provides insights into the influential capabilities of index participants with differing traits.

In settings without spillover, heterogeneous treatment effect (HTE) analysis has been adopted to identify the target individuals who have some predefined treatment effect (e.g., the largest treatment effect) in precision medicine \citep[e.g.,][]{khan2021precision} and differentiated marketing \citep[e.g.,][]{chen2021treatment}.
Current methods for HTE analysis can be classified into two categories \citep{Hu2023}.
In the first category, researchers use theories or expert knowledge to identify or construct subgroups where the treatment effect is heterogeneous. Then, statistical procedures, such as regression or semi-parametric methods, are adopted to reveal the pattern of interaction between the intervention and the subgroups 
\citep{cohen2013applied,  robins2000marginal}. Hence, this category is theory-driven and confirmatory.
The methods in this category mainly test the hypotheses of known subgroups instead of discovering unknown subgroups from data; these methods rely on clear prior hypotheses about which subgroups may be involved in the interactions. 
The other category  is data-driven and exploratory, and relies on machine learning methods developed for this purpose. For instance, tree-based recursive partitioning have been developed
\citep{chipman2010bart, athey2016recursive} to partition data into smaller subsets until a stopping criterion is met.
Recursive partitioning methods are a natural way to analyze a large number of features that have potentially complicated interactions. 
However, greedily learned trees can suffer from the lack of direct optimization of a global objective, often yielding suboptimal solutions \citep{kim2016hybrid}.

In this paper, we develop a confirmatory approach to be applied to ENRT designs identify subgroups of individuals who, if they are assigned to the intervention, would have the largest spillover effect on their network members.
Given the presence of multiple comparisons, we use a hypothesis testing method, the Multiple Comparison with the Best (MCB), to identify these best subgroups, while controlling for 
the family-wise error rate.
Traditional common-used multiple comparison methods, such as  Tukey's procedure \citep{tukey1991philosophy} and Dunnett's method \citep{dunn1961multiple}, 
are effective at controlling the overall significant level--the probability of rejecting at least one null hypothesis when all the null hypotheses are true, or equivalently, having at least one confidence interval that fails to include the true value.
These methods differ in how well they properly control the overall significant level and in their relative power. 
However, they primarily identify groups that are  significantly differently from one another. In many cases, the goal may extend beyond this to identifying subgroup(s) that is/are significantly better than others.
The multiple comparison with the best (MCB) approach
 \citep{hsu1996multiple} is a hypothesis testing method that focuses on identifying the superior choice among multiple alternatives while controlling for the risk of making false discoveries. It balances the desire to find the best option with the need to maintain the integrity of statistical inference. 
 \citet{zhu2015multiple} applied MCB in clinical problems to compare multiple treatment or prognostic groups with right-censored survival data, identifying the groups with the minimum risk. 
Similarly, \citet{artman2020power} and \citet{chao2022joint} used MCB to find optimal dynamic treatment regimens (DTR) with survival outcomes in the sequential, multiple assignment, randomized trial (SMART) setting and developed a power analysis framework.

In this paper, we contribute to the literature of HTE analysis by developing a regression method to evaluate the heterogeneity of spillover effects with respect to index-level characteristics in ENRT designs, extending the MCB testing approach to identify the subgroup of influencers with the largest spillover effects. In particular, we apply the MCB method to a linear mixed model 
for network members with index-level intervention accounting for clustering by network, we compute simultaneous confidence intervals that account formultiple comparisons of spillover effects of each subgroup with the best deriving group-specific critical values,  and we derive the overall p-value. 
A simulation study explores the size and the power of the MCB test.
Furthermore, we contribute to the literature on sample size calculations for causal inference under spillover \citep{Baird2018,jiang2023statistical} and under heterogeneity 
\citep{Yang2020,brookes2004subgroup}.
We derive the power function and methods for sample size calculations to design egocentric network-based studies to identify a subgroup of key influencers, that is, to detect a difference between the most influential subgroups and other subgroups. Specifically, we provide a procedure for determining the required minimal number of networks to detect differences in spillover effects with the desired power and significance level. We extend the standard MCB test to allow for the existence of multiple best subgroups.
%
Additionally, we compare the power and minimum sample size of the MCB test to an alternative test to reject the same null hypothesis of no heterogeneity of spillover effects.

The remainder of this paper is organized as follows. In Section \ref{sec:egonet}, we introduce the notation for the egocentric network-based randomized trial design, we define the casual estimands and provide the identifying assumptions.
In Section \ref{sec:regression}, we proposeregression-based estimators of 
heterogeneous spillover effects. 
The MCB method for identifying key influencers is presented in Section \ref{sec:identify_key}.
In Section \ref{sec:sample_size} we develop power and sample size calculations for our proposed method, and Section \ref{sec:simul} shows a simulation study.
In Section \ref{sec:data}, we apply the MCB method and sample size formulas to the STEP into Action study, a peer education intervention on HIV risk behaviors \citep{davey2011results}. Finally, we discuss our findings and potential future work in Section \ref{sec:discussion}.

\section{Heterogeneous Spillover Effects in Egocentric Network-based Studies}\label{sec:egonet}

\subsection{Setting and notation}
In a study of peer education interventions, a set of index participants
are asked to nominate a set of (social) network members, e.g., their sex partners, drug partners, or friends, as the intervention regions.
Using social network terminology, we refer to the network members of an index participant as \textit{egocentric network}, while we call  \textit{egonetwork} the set of an index participant and their nominated network members.

Let $\mathcal{N}$ be the study sample with $N = |\mathcal{N}|$ participants, indexed by
$ik$, 
where $k=1,\dots, K$ is the egonetwork indicator and $i=1,\dots, n_k$ is the participant indicator in egonetwork $k$.
Let $\mathcal{N}_k$ denote the sample of egonetwork $k$, composed of the index, denoted by $1k$, and the network members,denoted by $2k, \dots, n_kk$.
Using this notation, $\mathcal{N}=\{ik\}_{k=1, \dots, K; \,i=1, \ldots, n_k}$ represents our sample of units, and $\mathcal{N}_k=\{ik \in \mathcal{N}\}_{i=1, \dots, n_k}$ represents the subsample within each egonetwork $k$.

Denote by $Y_{ik}$ and $X_{ik}$ the observed outcome and covariate of unit $ik$, respectively.
In this paper, we consider $X_{ik}$ to be a categorical variable, i.e., $X_{ik} = h$ with $h=1,\dots, H$, representing the partition of the sample into pre-specified subgroups defined by individual socio-demographic characteristics measured at baseline. 

In an egocentric network-based randomized trial (ENRT), index participants are randomly assigned to a treatment (i.e., the training sessions), while all network members are left untreated. 
Let $Z_{ik} \in Z$ be the treatment status for subject $ik$, with $Z = \{0, 1\}$.  
Thus,
$Z_{1k}\in \{0,1\}$ and $Z_{ik}=0$ for $i>1$. 
We assume that index participants are randomly assigned to the intervention with probability $p$, with $0<p<1$, following a Bernoulli randomization, i.e., $Pr(Z_{1k}=1)=p$. Conversely, network members cannot receive the intervention, i.e., $Pr(Z_{ik}=1)=0$ for all $i>1$. However, network members can be indirectly exposed to the treatment received by their index participant.


\subsection{Identifying assumptions}


\subsubsection{Non-overlapping egonetworks}

In an ENRT, we only observe the connections of index participants $1k$, $k=1,\dots,K$, in the sample $\mathcal{N}$. 
However, the connections among network members $ik$, $i=2, \ldots, n_k$, 
are generally not observed. 
In an undirected network, the only observed connection for network members is the one with their index participant. Network members can potentially have unobserved connections with out-of-sample individuals and may be linked to other index participants $1k'$ with $k'\neq k$. 
Additionally, connections among network members in the same egonetwork are not observed.
Although by network transitivity it is likely that the peers of an index participant are also connected to each other, a fully connected egonetwork is not guaranteed and some pairs of network members of the same egonetwork may not be linked. Therefore, 
the observed egonetworks represent a partial network of the complete network of connections both among sampled participants and between  sampled participants and non-sampled individuals. 

For identification of causal effects
we assume that network members are only connected to one index participant in the sample and these index participants are not connected among themselves.
Let $\mathcal{N}_{ik}$ denote the set of individuals connected to unit $ik$. 
Let $\mathcal{N}^*_{ik}\subset \mathcal{N}$ be the network neighborhood of node $ik$ in sample $\mathcal{N}$, i.e., $\mathcal{N}^*_{ik}=\mathcal{N}_{ik} \cap \mathcal{N}$. Formally, we make the following assumption.

Assumption 1 (Non-overlapping Egonetworks): $(1k') \notin \mathcal{N}^*_{ik}$ for all $k'\neq k$.

\noindent $(1k') \notin \mathcal{N}^*_{1k}$ guarantees no connection between any two index participants. On the other hand, the assumption $(1k') \notin \mathcal{N}^*_{ik}$ for some $i\neq 1$ implies that the network member $ik$ can be connected to only one index, $1k$.
Note that assumption 1 allows for connections between two network members from different networks.
Assumption 1 is commonly ensured by protocol in ENRTs \citep{tobin2011step}.
 
\subsubsection{Neighborhood interference}

Under the potential outcome framework \citep{rubin2005causal}, we denote by $Y_{ik}(\z)$ the potential outcome of individual $i$ in egonetwork $k$ under the  treatment vector $\z$ in the population. 
Here, we relax the common no-interference assumption, allowing for the outcome of an individual $i$ in egonetwork $k$ to be affected not only by their own treatment $Z_{ik}$ but also by the treatment among the network neighbors of unit $ik$. 
This assumption, known as `neighborhood interference' in the causal inference literature, restricts interference to the network neighborhood \citep{forastiere2021identification,forastiere2022estimating, sussman2017elements}. 
Let  $\Z_{\mathcal{N}_{ik}}$ be the treatment vector in the neighborhood $\mathcal{N}_{ik}$ of unit $ik$ and $\Z_{-(ik, \mathcal{N}_{ik})}$ the treatment vector in the population excluding unit $ik$ and their neighborhood $\mathcal{N}_{ik}$, such that $\Z=[Z_{ik}, \Z_{\mathcal{N}_{ik}}, \Z_{-(ik, \mathcal{N}_{ik})}]$. Let $\z=[z_{ik}, \z_{\mathcal{N}_{ik}}, \z_{-(ik, \mathcal{N}_{ik})}]$ be a realization of the treatment vector.
Formally, the neighborhood interference assumption can be stated as follows.

Assumption 2 (Neighborhood interference): Given $\z$ and $\z'$ such that $z_{ik}=z'_{ik}$ and $\z_{\mathcal{N}_{ik}}=\z'_{\mathcal{N}_{ik}}$, then
	$Y_{ik}(z_{ik}, \z_{\mathcal{N}_{ik}}, \z_{-(ik, \mathcal{N}_{ik})})=Y_{ik}(z'_{ik}, \z'_{\mathcal{N}_{ik}}, \z'_{-(ik, \mathcal{N}_{ik})})$.

\noindent Under assumption 2, the potential outcome of unit $ik$ can be indexed as $Y_{ik}(z_{ik}, \z_{\mathcal{N}_{ik}})$. Note that assumption 2 only restricts interference to the neighborhood, but does not impose any specific mechanism of interference, allowing for heterogeneity in the way each neighbor's treatment may affect the outcome of unit $ik$.

Furthermore, we make the common consistency assumption \citep{pearl2010consistency,vanderweele2009concerning}, which implies that the observed outcome of unit $ik$, $Y_{ik}$, is equal to the potential outcome $Y_{ik}(Z_{ik}, \Z_{\mathcal{N}_{ik}})$ under the observed individual treatment $Z_{ik}$ and neighborhood treatment vector $\Z_{\mathcal{N}_{ik}}$.

\subsubsection{Randomization}

Let $R_{ik}$ be an indicator for whether unit $ik$ is an index participant, that is,  $R_{ik}=1$ if $ik$ is an index participant and 0 if it is a network member. Given our egocentric notation, it follows that $R_{1k}=1$ and $R_{ik}=0$ for all $i>1$. Under the randomization scheme of the ENRT design, the following assumption holds.

Assumption 3 (Randomization): $Y_{ik}(z_{ik}, z_{\mathcal{N}_{ik}})\independent Z_{ik}, \mathbf{Z}_{\mathcal{N}_{ik}}| R_{ik}=0$.

 This assumption states that 
potential outcomes for network members are independent of their individual treatment status and the treatment assignment within the neighborhood. Note that, given the ENRT design, $Z_{ik}=0$ for netork members $i>1$ and $\mathbf{Z}_{\mathcal{N}_{ik}}$ has all elements equal to 0 but the treatment of the index participant $Z_{1k}$, which can be 0 or 1. Given the randomization of $Z_{1k}$, assumption 3 is guaranteed.

\subsection{Causal estimands: heterogenous spillover effects}\label{sec:causal_effect}

Our causal estimand of interest is the 
spillover effect on the potential outcome of an untreated unit of the treatment received by a network neighbor, while the rest of the neighborhood remains untreated. We are interested in assessing the heterogeneity of such effect with respect to the individual characteristics of the treated neighbor. For this purpose, we define the following heterogeneous spillover effect:
\begin{equation}
\label{eq:ASPE}
\delta(h)=\E\Big[Y_{ik}\big(Z_{jk}=1, \Z_{(ik, \mathcal{N}_{ik}/jk)}=0\big)-Y_{ik}\big( Z_{jk}=0, \Z_{(ik, \mathcal{N}_{ik}/jk)}=0\big)| jk\in \mathcal{N}_{ik}, X_{jk}=h, R_{ik}=0\Big],
\end{equation}
where $\Z_{(ik, \mathcal{N}_{ik}/jk)}$ is composed of the individual treatment $Z_{ik}$ and the treatment vector of the neighborhood of unit $ik$, excluding unit $jk$. $\delta(h)$ is then the average spillover effect on an individual's outcome from the treatment of a neighbor in subgroup $h$ (i.e., with $X_{jk}=h$) while all other neighbors and including the individual are not treated.    
Note that the heterogeneous spillover effect is defined conditional on $R_{ik}=0$, that is, we define our estimand of interest only among network members in our ENRT design.
This is because network members and index participants and may exhibit different characteristics given the different way the two sub-samples are recruited. 
%
The expectation in Equation \eqref{eq:ASPE}
is taken with respect to the distribution of potential outcomes induced by the sampling of network members and by the selection of neighbor $jk$ among those with $X_{jk}=h$ to be assigned to treatment.



\subsection{Identification}

In the context of ENRTs, we aim to identify the  spillover effect $\delta(h)$ from the observed data. Given the characteristics of an ENRT design and under the assumptions of non-overlapping egonetworks and neighborhood interference, as well as the randomization of treatment among index participants and consistency, we can identify the spillover effects $\delta(h)$ as follows.
\vspace{-1em}
\begin{theorem}\label{thm:aspe}
 The spillover effect $\delta(h)$ from a network neighbor of subgroup $X_{jk}=h$ is identified by 
\begin{align}
\label{eq: identification}
\delta(h) = \E[Y_{ik}|Z_{1k}=1, X_{1k}=h, R_{ik}=0] - \E[Y_{ik}|Z_{1k}=0, X_{1k}=h, R_{ik}=0].
\end{align}   
\end{theorem}
\noindent The proof is provided in supplementary material S1. This theorem leverages the random assignment of treatments to index participants and the structure of the egocentric networks to isolate the spillover effects attributable to the treatment status of index participants in subgroup $h$.
According to Equation \eqref{eq: identification}, $\delta(h)$, i.e., the average spillover effect on an individual's outcome from the treatment of a neighbor in subgroup $h$, averaged over the conditional sampling distribution, is identified in the ENRT design by the mean difference in the observed outcomes between network members whose index participant is in subgroup $h$ and treated or in the subgroup $h$ but untreated.


\section{Regression-based Estimator}\label{sec:regression}

In this section, we provide a regression-based estimator to assess the heterogeneity of the spillover effect based on index-level characteristics.


\subsection{Statistical model}

Recall that $X_{ik}$ is a categorical variable representing a subgroup defined by the individual characteristics of the unit $ik$. Our goal is to investigate the heterogeneity of spillover effects from neighbors of different subgroups. 
Given the identification result in \eqref{eq: identification}, we consider the following model:
\begin{align}\label{eq:model}
Y_{ik} & = \sum_{h = 1}^{H} \zeta_{h} S_{kh} + \sum_{h = 1}^{H} \delta_{h} G_{ik}S_{kh} + u_k + \epsilon_{ik}, \quad k= 1,\dots, K,\  i = 2,\dots, n_k+1,
\end{align}
where $S_{kh}$ is the dummy variable of $X_{1k}$ with category $h$, i.e., $S_{kh} = I\{X_{1k} = h\}$, and $G_{ik}$ is 
the treatment status of the index participant in egonetwork $k$, i.e., $G_{ik}=Z_{1k}$ for $i>1$.
We assume that the residual error $\epsilon_{ik}$ and random network effect $\mu_k$ are independent, and they are normal distributed with zero means and variance $\sigma_e^2$ and $\sigma_u^2$, respectively. 
Note that the parameter $\delta_h$ identifies the spillover effect $\delta(h)$. In fact, we have
\vspace{-0.5cm}
\begin{align*}
    \delta(h) &= E[Y_{ik}|Z_{1k}=1, X_{1k}=h, R_{ik}=0] - E[Y_{ik}|Z_{1k}=0, X_{1k}=h, R_{ik}=0] \\
    &= E(Y_{ik}|Z_{1k}=1, G_{ik} = 1, S_{kh} = 1, R_{ik}=0) -
    E(Y_{ik}|Z_{1k}=0, G_{ik} = 0, S_{kh}=1, R_{ik}=0) \\
    &= \zeta_h + \delta_h - \zeta_h = \delta_h.
\end{align*}
\vspace{-1.3cm}

\noindent Thus, $\delta_{h}$ is interpreted as the average spillover effect for a network member whose index participant belongs to the subgroup $h$, with $h=1,\dots, H$.

Let $\bm{Y}_k = (Y_{1k},\dots, Y_{n_k,k})'$, $\bm{Z}_k = (Z_{1k},\dots,Z_{n_k,k})'$, $\bm{G}_k = (G_{1k},\dots,G_{n_k,k})'$, and $\bm{D}_k = (S_{k1}\bm{j}_k, \dots, S_{kH}\bm{j}_k$, $S_{k1}\bm{G}_k,\dots, S_{kH}\bm{G}_k)$, where $\bm{j}_k = (1,\dots,1)_{n_k}'$.
In our design, we assume that the outcomes of two individuals in different networks are independent, that is, \\
${\rm Cov}(Y_{ik},Y_{i'k}|\bm{Z}_k,\bm{G}_k) = 0$ for all $i\neq i'$, $i, i'>1$ , and all $k$.
Hence, ${\rm Var}(Y_{ik}) = \sigma^2  = \sigma_u^2 + \sigma^2_e$.
The intra-class correlation (ICC) between $Y_{ik}$ and $Y_{i'k}$, conditional on $\bm{Z}_k$ and $\bm{G}_k$, is defined as
$\rho_y = \frac{\sigma_u^2}{\sigma^2_u + \sigma^2_e}$.
As a result, ${\rm Var}(\bm{Y}_k|\bm{Z}_k,\bm{G}_k) = \sigma^2 \bm{V}_k$, where $\bm{V}_k = (1-\rho_y)\bm{I}_{n_k} + \rho_y \bm{J}_{n_k}$.
Let $\bm{\theta} = (\zeta_1,\dots, \zeta_H, \delta_{1},\dots,\delta_{H})$ be the vector of parameters,
the model \eqref{eq:model} can be rewritten as 
$$
\bm{Y}_k = \bm{D}_k \bm{\theta} + \mu_k + \bm{\epsilon}_k,\quad  k = 1,\dots, K.
$$

\subsection{The Generalized Estimating Equation (GEE) estimator}

Using Generalized Estimation Equations (GEE), the parameters of the model \eqref{eq:model}, \\$\bm{\theta}=(\zeta_{1},\ldots, \zeta_{H}, \delta_{1},\ldots, \delta_{H})$, can be estimated as 
\begin{align}\label{eq:var_cat}
    &\hat{\bm{\theta}} = \left(\sum_k \bm{D}^{'}_k \bm{V}_k^{-1} \bm{D}_k\right)^{-1} \left(\sum_k \bm{D}_k^{'} \bm{V}_k^{-1} \bm{Y}_k\right),
\end{align}
and $\text{Var}(\hat{\bm{\theta}}) = \sigma^2 \bm{U}^{-1}_{Ik}$, where $\bm{U}^{-1}_{Ik} = (\sum_k \bm{D}^{'}_k \bm{V}_k^{-1} \bm{D}_k)^{-1}$.
Under the standard regularity conditions \citep{zeger1988models}, as $n_k$ is fixed for all $k=1,\dots,K$, and $K \rightarrow \infty$, $\sqrt{K}(\hat{\bm{\theta}}-\bm{\theta})$ is asymptotically normally distributed as $N(0, \bm{\Sigma}_I)$, where $\bm{\Sigma}_I = \lim_{K \rightarrow \infty} \sigma^2 (\U_{Ik}/K )^{-1}=\sigma^2 \U_I^{-1}$ with $\U_I = \lim_{K \rightarrow \infty} \frac{1}{K} \U_{Ik} $. 

\begin{lemma}\label{lem:var_cat}
Suppose $g_h = \sum_{k=1}^K S_{kh}/K$ be the proportion for category $h$, $h = 1,\dots, H$. Let $b_k = n_k(c+d_kn_k)$ where $c =\frac{1}{1-\rho_y} $ and $d_k = -\frac{\rho_y}{(1-\rho_y)(1+n_k\rho_y)}$, and we define $\bar{b} = \lim_{K\rightarrow \infty} \sum_{k=1}^K b_k/K$. For simplicity, we assume all networks have the same size, i.e., $n_k=n$,
then $\bar{b} = \frac{n}{(1-\rho)(1+n\rho)}$.
%
Let $\bm{\delta} = (\delta_1,\dots, \delta_{H})$ be the parameters of interest, with corresponding estimates $\hat{\bm{\delta}} = (\hat{\delta_1},\dots,\hat{\delta}_{H})$, then
\begin{align*}
{\rm Var}(\hat{\bm{\delta}}) = \bm{\Sigma}_{\hat{\bm{\delta}}} 
= \frac{\sigma^2}{1-p} 
\begin{psmallmatrix}
\frac{1}{\bar{b} pg_1} & & \\
& \ddots & \\
& & \frac{1}{\bar{b} pg_H}
\end{psmallmatrix}.
\end{align*}
\end{lemma}


\section{Hypothesis Tests for Spillover Effect Heterogeneity and for the Identification of Key Influencers} \label{sec:identify_key}

Key influencers with the largest spillover effects emerge only when there is heterogeneity in the spillover effect across individuals or subgroups. Identifying such influencers requires not only estimate spillover effects but also assessing the differences in these effects across the network. 
For the testing of heterogeneity, the null hypothesis assumes that 
there is no best subgroup (category) among the competing options, that is, all the average spillover effect of an individual's treatment on neighbors' outcomes are the same across treated subgroups. On the contrary,  the alternative hypothesis indicates that there is a best subgroup, that is, there is a subgroup of individuals who, if treated, will have the highest average outcome among their network neighbors.
A common approach to test for this heterogeneity is the Wald \citep{wald1943tests} test, which evaluates whether variations in spillover effects are statistically significant different. In Section \ref{sec:wald} we first present the Wald test for the heterogeneity in  spillover effects. Then in Section \ref{sec:mcb} we introduce the MCB procedure to simultaneously identify the subgroup(s) with the largest  spillover effect and test for heterogeneity.

\subsection{Wald tests}
\label{sec:wald}
    

For testing the heterogeneity of spillover effects, the hypotheses are:
$
    H_0: \delta_1=\dots= \delta_{H}, \;
    H_1: \delta_h \neq \delta_j \quad \text{for some $h\neq j$},
$
which are equivalent to
$
H_0: \delta_1 -\delta_H = \delta_2 -\delta_H = \dots = \delta_{H-1}-\delta_H = 0,\;
H_1: (\delta_1-\delta_H, \delta_2-\delta_H, \dots,\delta_{H-1}-\delta_H) \neq \bm{0}.
$
 Let $\bm{\delta}_{-H} =(\delta_1-\delta_H,\delta_2-\delta_H,\dots, \delta_{H-1}-\delta_H)$ denote the vector of difference, with the corresponding estimator $\hat{\bm{\delta}}_{-H}$. The Wald test statistic is then
$
  W(\hat{\bm{\delta}}_{-H}) = \hat{\bm{\delta}}_{-H}^T [\text{Var}(\hat{\bm{\delta}}_{-H})]^{-1}\hat{\bm{\delta}}_{-H},
$   
which asymptotically follows a $\chi^2$ distribution as $K\rightarrow \infty$. 



\subsection{Multiple comparisons with the best (MCB)}

Identifying which subgroup(s) has the largest spillover effect may be of interest, both to investigate spillover heterogeneity and to select the optimal set of participants to be intervented.
The MCB procedure permits the identification of a confidence set of best candidates which cannot be statistically
distinguished from the unknown true best for the given data while adjusting for multiple comparisons \citep{hsu1984constrained, hsu1996multiple}. 
Specifically, 
the MCB simultaneously tests the difference between the spillover effect of each subgroup $h$ and that of the best of the others. Without loss of generality, we assume that a larger effect indicates a better group. That is, when we are interested in identifying the subgroup(s) with the highest spillover effect, 
for each subgroup $h$, the MCB tests the hypotheses
$
H_0^h:\quad \delta_h \geq \max_{j \neq h} \delta_j, \;
H_{1}^h: \quad \delta_h < \max_{j\neq h}\delta_j, 
$

where the subgroup-specific null assumes that subgroup $h$ is the best subgroup or not worse than the best one. As such, we have that when for all subgroups the individual null hypothesis is not rejected, then the null hypothesis of no heterogeneity $H_0: \delta_1=\dots= \delta_{H}$ cannot be rejected, i.e., 
$\cap_{h=1}^H H^h_0=H_0$. On the other hand, when at least one individual null $H_0^h$ is rejected, then the overall null of no heterogeneity $H_0$ is also rejected. MCB tells us with a specified level of confidence, which groups may be the best, and it provides upper and lower bounds on the deviations of all the groups from the best of others.
Denote by $B_0=\{h: H_0^h \,\,\text{is true}\}$
the set of indices of true best subgroups, and by $B_1=\{h: H_0^h \,\,\text{is not true}\}$  the set of indices of non-best subgroups.

Group $h$ is considered statistically indistinguishable from the true best at a significance level $\alpha$  if and only if for all $j \neq h$,
$
  \frac{\hat{\delta}_h - \hat{\delta}_j}{\sqrt{\text{Var}(\hat{\delta}_h - \hat{\delta}_j)}} \geq - c^{h}_{\alpha},  
$
where $c^h_{\alpha}$ is the critical value, which is the solution to 
\begin{align}\label{eq:critical}
	P(\max_{j=1,\dots,H-1} z_j \leq c^h_{\alpha}) = 1-\alpha,
\end{align}	
where $(z_1, \dots, z_{H-1})$ is an random vector distributed as multivariate t-distribution $t_{\nu}(\bm{R}_h)$, $\nu$ is the degree of freedom for the $\chi^2$ distribution of  $\hat{\sigma}/\sigma$, where $\hat{\sigma}^2$ is the estimated residual variance 
, and $\bm{R}_{h}$ is the correlation matrix between $\hat{\delta}_j$ and $\hat{\delta}_h$. 
The critical value $c^h_{\alpha}$ depends on $h$, because the correlation matrix $\bm{R}_h$ depends on $h$.
One way to calculate the critical value $c^h_{\alpha}$ is using simulations, with $H$ distinct simulations, one for each value of $h$. When $H$ is large, this will be extremely time-consuming both because the number of simulations must be large, and because each of the individual simulations is complicated (involving $H-1$ dimensional probability calculation).
Therefore, to compute the critical values we borrow the idea of \citet{hsu1996multiple}, who proposed a computationally efficient procedure.
\vspace{-1em}
\begin{lemma}\label{lem:critival2}
Let $\hat{\bm{\delta}}_{-h} = (\hat{\delta}_j - \hat{\delta}_h)$, $\forall j\neq h$.
Under the iid normal error assumption of our model, $\hat{\bm{\delta}}_{-h}$ is multivariate normally distributed and  $\nu \hat{\sigma}^2/\sigma^2$ has a $\chi^2$ distribution.
Define $\bm{R}_{-h}$ as the correlation matrix of $\hat{\bm{\delta}}_{-h}$, i.e.,
$
	\bm{R}_{-h} = \text{diag}(1-(\lambda_1^h)^2, \dots, 1-(\lambda_{H-1}^h)^2) + (\lambda_1^h, \dots, \lambda_{H-1}^h)^T (\lambda_1^h, \dots, \lambda_{H-1}^h)
$.
	Then from \eqref{eq:critical} the critical value $c^h_{\alpha}$ can be written as the solution to 
\begin{align*}
	\int_0^{\infty}\int_{-\infty}^{\infty} \prod_{j=1}^{H-1} \Phi(\frac{\lambda_j^h z + c^h_{\alpha} u}{\sqrt{1-(\lambda_j^h)^2}}) \phi(z)\gamma(u) du = 1-\alpha, 
\end{align*}
where $\Phi(\cdot)$ and $\phi(\cdot)$ are the cdf and pdf of the standard normal distribution, respectively. $\gamma(\cdot)$ is the density function of $\hat{\sigma}/\sigma$.
\end{lemma}

\vspace{-2em}
\begin{remark}
There might exist multiple solutions of $(\lambda^h_1, \dots, \lambda^h_{H-1})$ that $\rho_{ij}^h = \lambda_i^h \lambda_j^h$ for all $i,j \leq H-1$, but it does not affect the choice of $c^h_{\alpha}$ (see the proof of Lemma \ref{lem:critival2}).
\end{remark}
\vspace{-1em}


Denote the variance of $\hat{\delta}_h - \hat{\delta}_j$ by $\sigma^2 v_{jh}$,
and define the estimated set of the best subgroups as $
\widehat{B_0} := \{h: \hat{\delta}_h \geq \max_{j\neq h}[\hat{\delta}_j - c^h_{\alpha} \sigma\sqrt{v_{jh}}] \}$.
It is straightforward to show that under the overall null (i.e., when there is no heterogeneity), the probability that the set of best includes all subgroups is at least $1-\alpha$:
$
P(h \in \widehat{B_0}| \delta_1 = \delta_2=\dots=\delta_H) \geq 1- \alpha \quad \text{for all } h = 1, \dots,H.
$
\noindent This coincides with the probability that the overall null $H_0$ is not rejected when it is in fact true, i.e., $P(\text{fail to reject}\,\, H_0| H_0\  \text{is true}\}$.
This means that by identifying the best subgroup(s)  $
\widehat{B_0}$ using critical values, we ensure that the overall Type I error is controlled and lower than $\alpha$.


\subsubsection{Simultaneous confidence intervals}

Given the critical values $c_{\alpha}^h$, $h=1, \dots, H$, in the previous section, we can now adjust  confidence intervals of $\hat{\delta}_h$ for multiple comparisons.
Suppose 
the $1-\alpha$ confidence interval for $\delta_h-\delta_j$ is $[L_{hj}, U_{hj}]$, i.e., $P(\delta_h-\delta_j \in [L_{hj}, U_{hj}]) = 1-\alpha$, then the probability that all the intervals cover the corresponding true values is $P(\delta_h-\delta_j \in [L_{hj}, U_{hj}]\ \text{for all } h,j, h\neq j) < 1- \alpha$.
Hence, without adjustment for multiplicity, the rate of making incorrect decisions may be unacceptably high, and we thus derive simultaneously confidence intervals to maintain the overall specified confidence level. 

Following to \citet{hsu1984constrained}, the  simultaneous confidence interval for $\delta_h - \max_{j\neq h} \delta_j$ is defined as follows:
\begin{equation}
\label{eq:SCI}
\begin{aligned}
U_h = \max\{0, \min_{j\neq h} (\hat{\delta}_h - \hat{\delta}_j + c^h_{\alpha} \hat{\sigma} \sqrt{v_{jh}}) \},\; S = \{h : U_h >0\}, \;
L_h = \min\{0, \min_{j\in S, j\neq h} (\hat{\delta}_h - \hat{\delta}_j - c^h_{\alpha} \hat{\sigma} \sqrt{v_{jh}}) \}.	
\end{aligned}
\end{equation}
Then the confidence intervals are $\delta_h - \max_{j\neq h}\delta_j \in [L_h, U_h]$ for $h=1,\dots, H$. 

A confidence interval for $\delta_h - \max_{j\neq h} \delta_j$ whose lower bound is 0 indicates the $h$th group is the best, and a confidence interval for $\delta_h -\max_{j\neq h} \delta_j$ whose upper bound is 0 indicates the $h$th group is not the best. A lower bound of 0 means the $h$th group is one of the best, and a lower bound close to 0 indicates the $h$th group is close to the best (assuming a larger $\delta_h$ implies a better group). The lower bounds measure how much the groups identified not to be the best are worse than the true best. 
\vspace{-1em}
\begin{theorem}\label{thm:coverage}
As $K\rightarrow \infty$, under the regularity conditions of the GEE estimator $\hat{\bm{\delta}}$, the intervals $[L_h, U_h]$, $h=1,\dots,H$, defined as in Equation \eqref{eq:SCI}, are a set of $100(1-\alpha)\%$ asymptotic simultaneous confidence intervals for $\delta_h -\max_{j\neq h} \delta_j$: $P(\delta_h -\max_{j\neq h} \delta_j \in [L_h, U_h] \ \forall h) \geq 1-\alpha$. Furthermore, when there is a unique critical value $c^{\alpha}_h$, the asymptotic coverage of the simultaneous confidence interval is exactly $1-\alpha$: $P(\delta_h -\min_{j\neq h} \delta_j \in [L_h, U_h] \ \forall h) = 1-\alpha.	$,  otherwise the asymptotic coverage is strictly greater than $1-\alpha$.
\end{theorem}
\vspace{-1em}

\noindent The proof is given in supplementary material S1. This theorem guarantees that the simultaneous confidence intervals provided by the MCB procedure maintain at least $1 - \alpha$ probability of correctly covering each true parameter difference $\delta_h - \max_{j\neq h}{\delta}_j$ across all groups $h$.

\vspace{-1em}
\begin{remark}
Suppose that a smaller treatment effect implies a better treatment, then the parameters of interest become $\delta_h - \min_{j\neq h} \delta_j, h=1,\dots, H$. Using the same definitions of the critical value $c^h_{\alpha}$ in \eqref{eq:critical}, but the simultaneous confidence intervals becomes
\begin{align*}
L_h = \min\{0, \max_{j\neq h} (\hat{\delta}_h - \hat{\delta}_j - c^h_{\alpha} \hat{\sigma} \sqrt{v_{jh}}) \},\; S = \{h : L_h < 0\}, \;
U_h = \max\{0, \max_{j\in S, j\neq h} (\hat{\delta}_h - \hat{\delta}_j + c^h_{\alpha} \hat{\sigma} \sqrt{v_{jh}}) \}.	
\end{align*}
Similarly to the previous case,  the coverage rate of confidence intervals still has the property that
$
P(\delta_h -\min_{j\neq h} \delta_j \in [L_h, U_h] \ \forall h) \geq 1-\alpha.	
$
\end{remark}

\subsubsection{P-value in MCB}\label{sec:p-value}

We define the MCB test statistics for subgroup $h$ as $T_h = \hat{\delta}_h - \max_{j\neq h}\hat{\delta}_j$ for $h=1,\dots, H$. Let $t_h$ be the observed test statistics from the distribution of $T_h$.
Suppose that the overall null hypothesis $H_0: \delta_1=\dots = \delta_H$ is true, the overall p-value can be defined as 
\begin{align*}
\text{p-value}_{\text{overall}} 
& = P(T_h \geq t_h \ \text{for all $h=1,\dots, H$}|H_0) \\
&= P(\hat{\delta}_h - \max_{j\neq h}\hat{\delta}_j \geq t_h \ \text{for all $h=1,\dots, H$}|H_0) \\
&= P(\frac{\hat{\delta}_h - \max_{j\neq h} \hat{\delta}_j}{\sigma \sqrt{v_{jh}}} \frac{\sigma}{\hat{\sigma}} \geq \frac{t_h}{\hat{\sigma}\sqrt{v_{jh}}} \ \text{for all $h=1,\dots, H$}|H_0) \\
&= 1 -  \int_{0}^{\infty}\int^{\infty}_{-\infty} \prod_{j\neq h, h=1,\dots H} \Phi\left(\frac{\lambda_j^h z + s t_h/(\hat{\sigma} \sqrt{v_{jh}})}{\sqrt{1-\lambda^h_j}}\right)\phi(x) \gamma(s) dz ds.
\end{align*}
\noindent The last line follows from the derived distribution of $\frac{\hat{\delta}_h - \max_{j\neq h} \hat{\delta}_j}{\sigma \sqrt{v_{jh}}}$ (see the proof of Lemma \ref{lem:critival2} in supplementary material S1).

On the other hand, for the p-value corresponding to the pairwise comparison with the best the marginal distribution of $\hat{\delta}_h - \max_{j\neq h} \delta_j$ is a non-central t distribution. However, pairwise p-values can be adjusted by using the conservative Bonferroni correction method which multiplies the raw p-values by the number of tests \citep{dunn1961multiple}. The adjustment approach overcorrects according to the false positive rate (family-wise error rate). Another very common method is to use Benjamini-Hochberg \citep{benjamini1995controlling} (BH) which is more powerful and controls the false discovery rate (FDR). 

\section{Power and Sample Size} \label{sec:sample_size}
Here, we derive the power and sample size formulas to detect an heterogeneity in the spillover effect and to correctly identify the best subgroup(s).
For simplicity, we assume that all index participants have the same number of network members, denoted as $n_k = n$ for $k=1,...,K$.
\subsection{Wald test}

Recall that under the null hypothesis $H_0$, the Wald test statistic follows a central $\chi^2 (p)$ distribution with $p$ being the degree of freedom.
While under $H_1: \bm{\delta}_{-H} = (\delta_1 -\delta_H, \delta_2-\delta_H, \dots = \delta_{H-1}-\delta_H) \neq \bm{0}$, it follows a $\chi^2(H-1,\vartheta)$ distribution, where $H-1$ is the degree of freedom, and $\vartheta$ is the non-central parameter,  defined as $\bm{\delta}_{-H}^T [\text{Var}(\hat{\bm{\delta}}_{-H})]^{-1} \bm{\delta}_{-H}$.
Therefore, the power function of the Wald test is defined as
$
    \text{Power}_{\rm Wald} = \int_{\chi_{1-\alpha}^2(H-1)}^{\infty} f(x;H-1,\vartheta)\ dx,
$

where $f(x;H-1,\vartheta)$ is the pdf of $\chi^2(H-1,\vartheta)$.

Under a given power size $1-\beta$, the sample size required for testing the heterogeneity of spillover effects can be calculated by
\begin{align}\label{eq:multiX}
   \int_{\chi_{1-\alpha}^2(p)}^{\infty} f(x;p,\vartheta(K))\ dx \geq 1-\beta.
\end{align}
As $\int_{\chi_{1-\alpha}^2(p)}^{\infty} f(x;p,\vartheta(K))\ dx$ is an increasing function with respect to sample size $K$, we can find the minimal $K$ such that \eqref{eq:multiX} holds through numerical methods.

\subsection{MCB test} \label{sec:mcb}

\citet{hsu1996multiple} defined the power of MCB test as the probability of coverage rate and narrow confidence intervals when at least one individual difference is non-zero. However, when there are multiple true best groups, their corresponding  individual difference is 0 (see Example 2 and 3 below). Hence, we extend the definition and expression of the power for the scenario with multiple best groups. 
In general, power is defined as the probability of rejecting the null when the null is false and not rejecting it when it is true: 
\begin{align}\label{eq:power_dnf}
	\text{Power}_{\text{MCB}} &= P({\rm Reject}\ H_0^h|H_1^h\ \text{for all}\ h \in B_1\  \& \ {\rm Accept}\ H_0^{h'}|H_0^{h'} \ \text{for all}\ h' \in B_0 ) 
\end{align}

Following \citet{hsu1996multiple}, we calculate \eqref{eq:power_dnf} both based on the consideration of coverage of confidence intervals and their width. 
For a non-best subgroup $h \in B_1$, one way to reject $H_0^h$ is to construct a sufficiently narrow confidence interval for $\delta_h -\max_{j\neq h}\delta_j$, while still covering the true value. That is, $\delta_h - \max_{j\neq h}\delta_j > \hat{\delta}_h - \max_{j\neq h}\hat{\delta}_j -c_{\alpha}^h$ and $c_{\alpha}^h\hat{\sigma}\sqrt{v_{jh}} \leq |\delta_h -\max_{j\neq h} \delta_j|$. For a best subgroup $h' \in B_0$, $H_0^{h'}$ cannot be rejected as long as the confidence interval covers the true value, regardless of the width. That is, $\delta_{h'} - \max_{j\neq {h'}}\delta_j \leq \hat{\delta}_{h'} - \max_{j\neq {h'}}\hat{\delta}_j +c^{h'}_{\alpha}\hat{\sigma}\sqrt{v_{h'j}}$.
Hence, the MCB power in \eqref{eq:power_dnf} can be writen as 
{
\begin{equation}\label{eq:power_cal}
\begin{aligned}
	\text{Power}_{\text{MCB}} 
	&= P\{(\delta_h - \max_{j\neq h}\delta_j > \hat{\delta}_h - \max_{j\neq h}\hat{\delta}_j -c_{\alpha}^h\hat{\sigma}\sqrt{v_{jh}} \ \text{and } c_{\alpha}^h\hat{\sigma}\sqrt{v_{jh}} \leq \max_{j\neq h} \delta_j - \delta_h \\
&\quad \text{for all h}\in B_1, \;   
	  \delta_h - \max_{j\neq h}\delta_j \leq \hat{\delta}_h - \max_{j\neq h}\hat{\delta}_j -c^h_{\alpha}\hat{\sigma}\sqrt{v_{hj}} \ \text{for all h} \in B_0\} \\
	& \geq \int_0^{\infty} \int_{\-\infty}^{\infty} \prod_{h \in B_0} \left[1- \Phi(\frac{\lambda_h z + c^h_{\alpha} u}{\sqrt{1 - \lambda_h^2}})\right] r(u) d\Phi(z)du \times \\
	& \quad \int_0^{u^*} \int_{\-\infty}^{\infty} \prod_{h \in B_1} \Phi(\frac{\lambda_h z + c^h_{\alpha} u}{\sqrt{1 - \lambda_h^2}}) r(u) d\Phi(z)du.
\end{aligned}
\end{equation}
}
where $u^* = \min_h \{\max_{j\neq h} (\delta_j -\delta_h)/(c^h_{\alpha} \sigma \sqrt{v_{hj}})\}$.
It can be seen that the power is approximated by the probability of confidence coverage when the individual nulls are true, and the probability of confidence coverage and narrow confidence intervals when the individual nulls are not true.

To better understand the MCB power, we consider the following three distinct examples:
\begin{example}
Let  $\delta_1 = \delta_2 = \delta_3 = 0$ and $\delta_4 = 0.5$. Then $\bm{\delta}_{-H} = (-0.5, -0.5, -0.5)$. In this example, there is only one single best subgroup.
Then \\
$
{\rm Power}_{\rm MCB} = \int_0^{u^*} \int_{\-\infty}^{\infty} \prod_{h \in \{1,2,3\}} \Phi(\frac{\lambda_h z + c^h_{\alpha} u}{\sqrt{1 - \lambda_h^2}}) r(u) d\Phi(z)du.
$

\end{example}

\begin{example}
Let  $\delta_1 = \delta_2 = 0$ and $\delta_3 = \delta_4 = 0.5$. Then $\bm{\delta}_{-H} = (-0.5, -0.5, 0)$ In this example, there are multiple best subgroups.
Then 
\small
$$
{\rm Power}_{\rm MCB} = \int_0^{\infty} \int_{\-\infty}^{\infty} \prod_{h=3} \left[1- \Phi(\frac{\lambda_h z + c^h_{\alpha} u}{\sqrt{1 - \lambda_h^2}})\right] r(u) d\Phi(z)du \times  \int_0^{u^*} \int_{\-\infty}^{\infty} \prod_{h \in \{1,2\}} \Phi(\frac{\lambda_h z + c^h_{\alpha} u}{\sqrt{1 - \lambda_h^2}}) r(u) d\Phi(z)du.
$$
\end{example}

\begin{example}
Let  $\delta_1 = 0$ and $\delta_2 = \delta_3 = \delta_4 = 0.5$. Then $\bm{\delta}_{-H} = (-0.5, 0, 0)$. In this example, there are multiple best subgroups.
Then 
\small
$$
{\rm Power}_{\rm MCB} = \int_0^{\infty} \int_{\-\infty}^{\infty} \prod_{h\in \{2,3\}} \left[1- \Phi(\frac{\lambda_h z + c^h_{\alpha} u}{\sqrt{1 - \lambda_h^2}})\right] r(u) d\Phi(z)du \times  \int_0^{u^*} \int_{\-\infty}^{\infty} \prod_{h =1} \Phi(\frac{\lambda_h z + c^h_{\alpha} u}{\sqrt{1 - \lambda_h^2}}) r(u) d\Phi(z)du.
$$
\end{example}

\begin{remark}
On the connection between the power function and narrowness of confidence intervals. Consider Example 2, for testing $H_{0}^3: \delta_3 - \max_{j\neq 3}\delta_j \geq 0$, that is, group 3 is one of the best.  
The upper bound for $\hat{\delta}_1 - \max_{j\neq 1}\hat{\delta}_j$ is always positive whatever the narrowness of the confidence interval, because its distribution is centered at a non-negative value.
Thus, we cannot reject the null assumption that group 3 is the best.
However, when testing the null for subgroup 1, with  
$\delta_1-\delta_4 = -0.5$, the distribution of $\hat{\delta}_1 - \max_{j\neq 1}\hat{\delta}_j$ is centered at $\delta_1-\max_{j\neq 1}\delta_j = -0.5$, and under the narrowness condition, the upper bound will be 0, and thus we reject the null that no any group is the best.
Therefore, coverage and narrowness conditions can imply the rejection of the null when it is not true.
\end{remark}


In general, given confidence level $1-\alpha$, the sample size is calculated such that, with a pre-specified probability $1-\beta\  (<1-\alpha)$, the confidence intervals will cover the true parameter and be sufficiently narrow.
The power function defined in \eqref{eq:power_cal} involves a double integral of complex functions, with the sample size influencing $\lambda_h$ and  $c^h_{\alpha}$. As a result, deriving a closed-form solution for the sample size is highly challenging.
To address this, in our implementation the power is given graphically as a function of the sample size, which allows for the rapid determination of the appropriate sample size by simple inspection.

\subsection{Comparison of Wald and MCB tests}\label{sec:comp}

Although both the Wald and MCB procedures test for heterogeneity of spillover effects, MCB further identifies the best subgroup(s) based on the simultaneous comparisons for each subgroup with adjustment for multiple comparisons.
Intuitively, the Wald test is expected to be more powerful since it only involves a single test of multiple parameters.

Here, we compare the efficiency of the two tests and investigate how design input parameters affect the power. In particular, 
we investigate the role of ICC $\rho_y$, the number of network members $n$, and the effect sizes $\Delta_{\delta}$ in heterogeneity of spillover effects by simulations. 
For fixed $\sigma^2=1$ and $p=0.5$,  we let $\rho_{Y}$, vary  from 0 to 1, $n \in \{2,3,5,10\}$, and calculate the number of required egonetworks $K$ for $\alpha = 0.05$ and $\beta=0.1$. 
Here we consider the covariate $X_{1k}$ to be categorical with H categories, where we let $H=3,4, 5$ and 6. 
In addition, we consider the structure of alternative hypothesis as in Example 1 (unique best) and Example 3 (multiple best) in Section \ref{sec:mcb}.

\begin{figure}[ht!]
    \centering
    \begin{subfigure}[t]{0.55\textwidth}
        \centering
        \includegraphics[width=\textwidth]{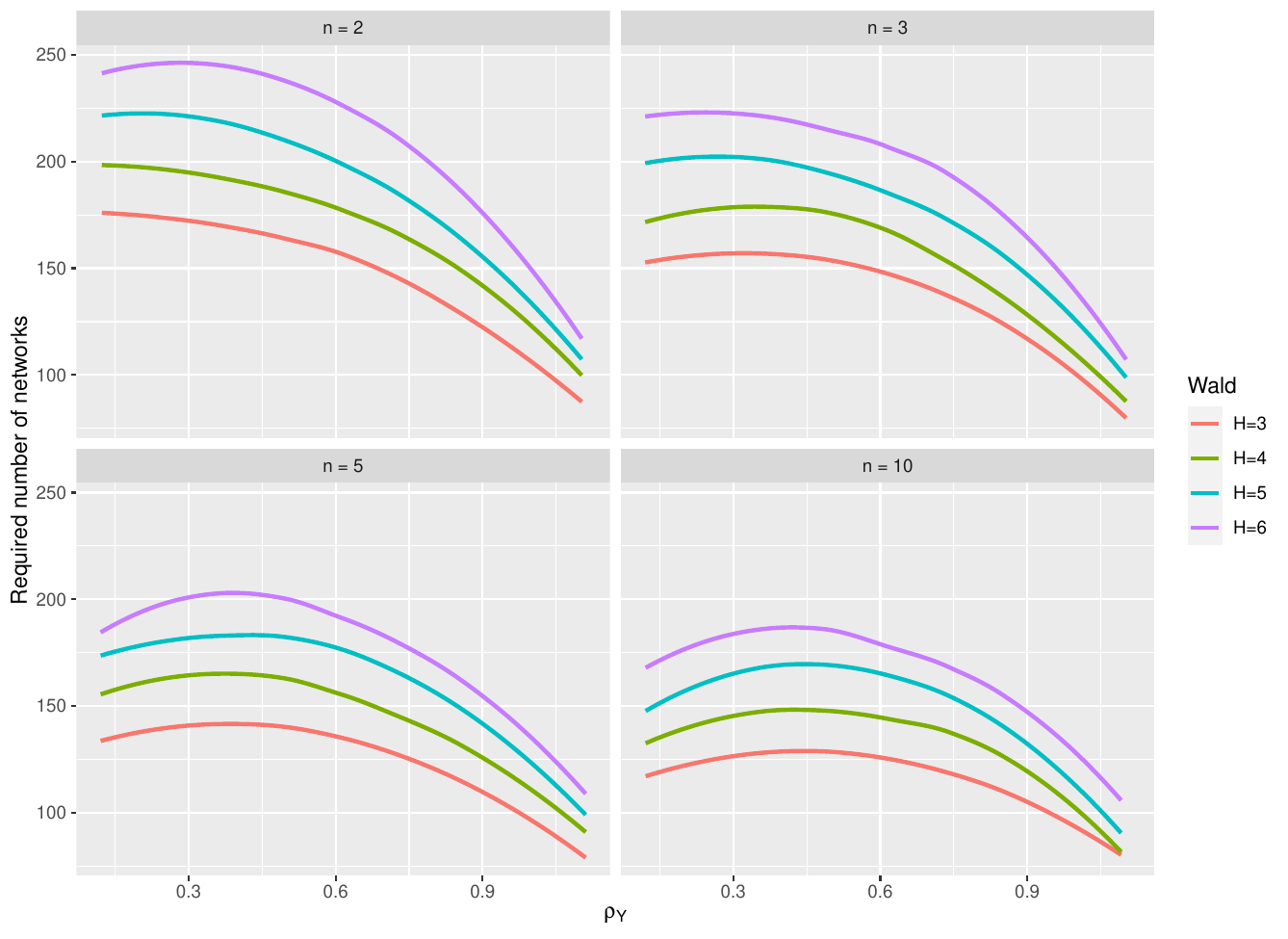}
        \caption{Wald test}
    \end{subfigure}%
~
    \begin{subfigure}[t]{0.55\textwidth}
        \centering
        \includegraphics[width=\textwidth]{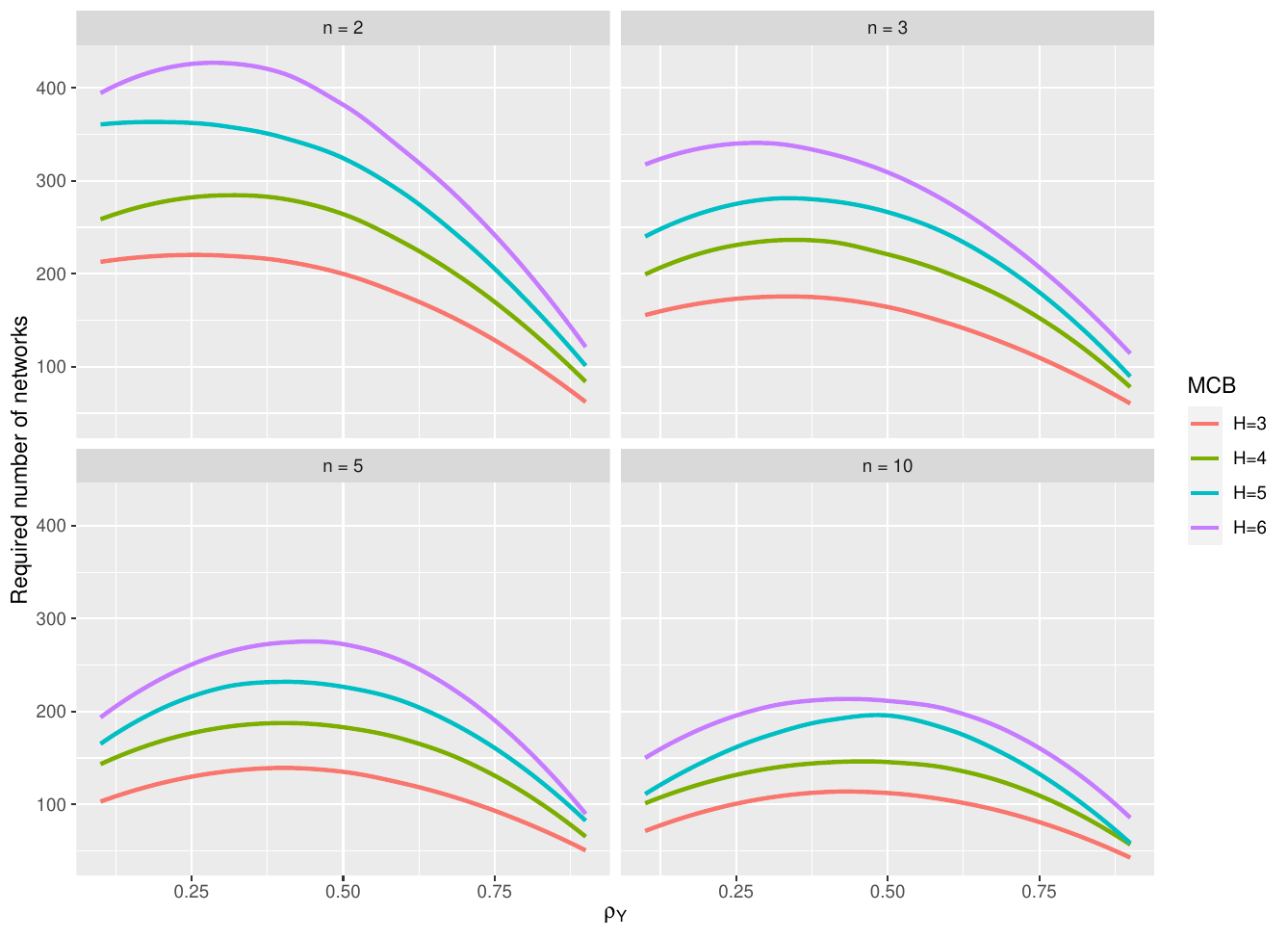}
        \caption{MCB test}
    \end{subfigure}
    \caption{Minimum required number of egonetworks for the Wald test (left) and the MCB test (right) for testing the heterogeneity of spillover effects when $H=3,4,5,6$ for Example 1 in which $\bm{\delta}_{-H}=c(-0.5,-0.5,-0.5)$. }
        \label{fig:wald_mcb_rep}
\end{figure}

\begin{figure}[ht!]
    \centering
    \begin{subfigure}[t]{0.55\textwidth}
        \centering
        \includegraphics[width=\textwidth]{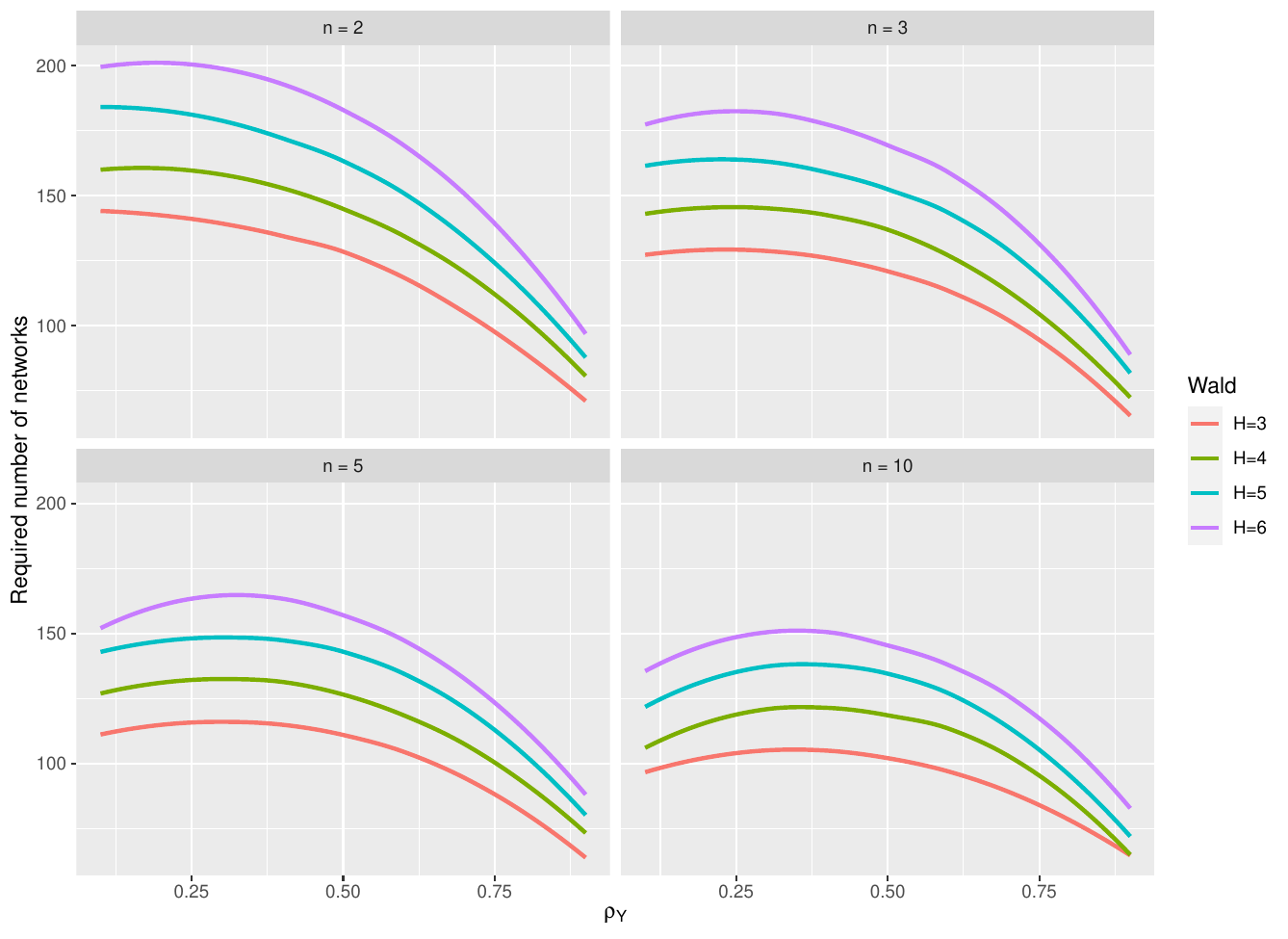}
        \caption{Wald test}
    \end{subfigure}%
 ~
    \begin{subfigure}[t]{0.55\textwidth}
        \centering
        \includegraphics[width=\textwidth]{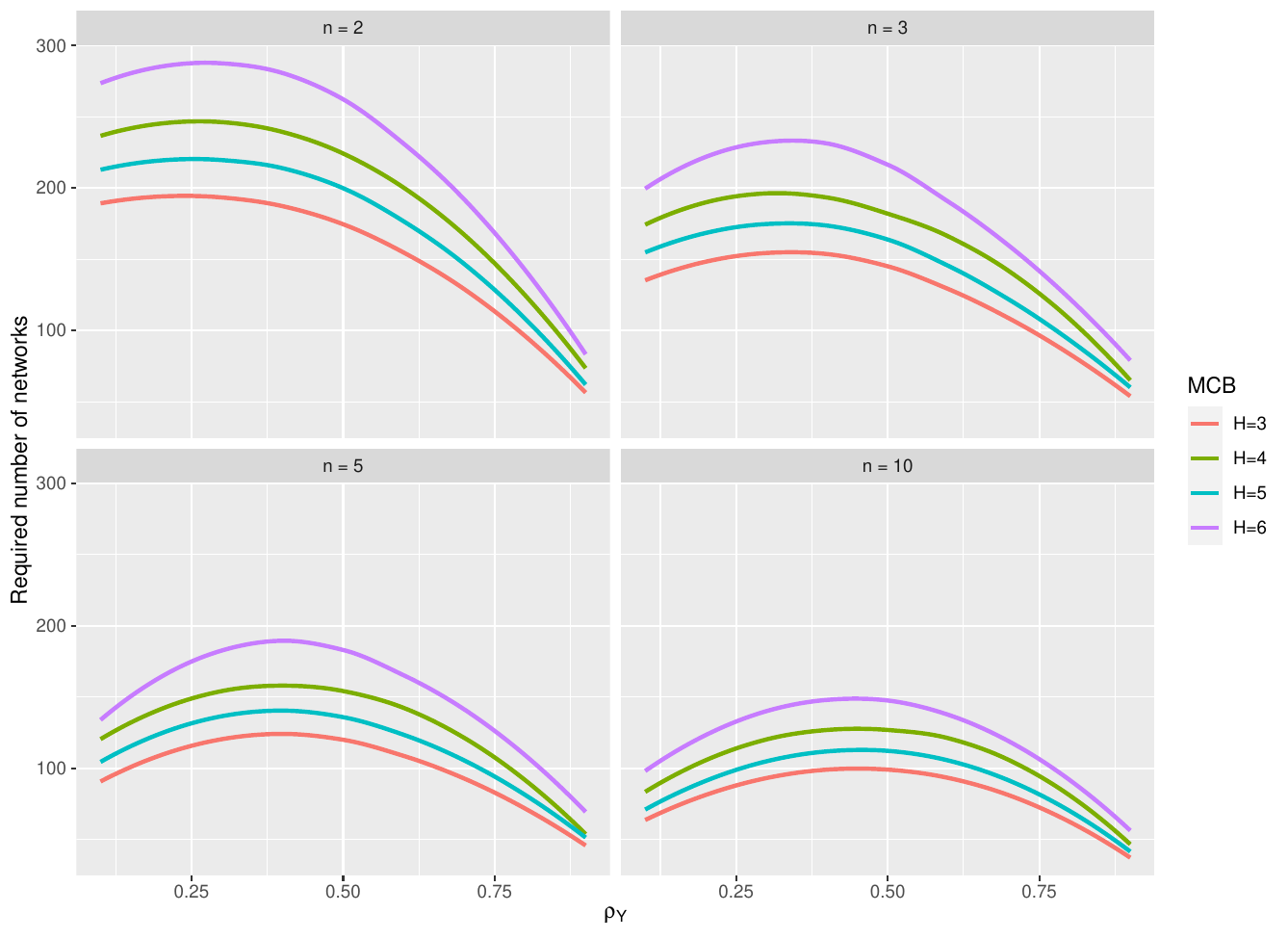}
        \caption{MCB test}
    \end{subfigure}
    \caption{Minimum required number of egonetworks for the Wald test (left) and the MCB test (right) for testing the heterogeneity of spillover effects when $H=3, 4, 5, 6$ for Example 3 in which $\bm{\delta}_{-H}=c(-0.5,0,0)$.}
        \label{fig:wald_mcb_nrep}
\end{figure}

Figure \ref{fig:wald_mcb_rep} , corresponding to Example 1,  and Figure \ref{fig:wald_mcb_nrep}, corresponding to Example 3, show the relationship between the required number of egonetworks and the parameters, $\rho_y$ and the number of social network members in each egonetwork $n$. It is seen that for a given power, a larger $\rho_y$ tends to require, after an initial increase, a lower $K$. This means that the stronger the network members' outcome distribution resembles  that of their index participant, the fewer number of networks $K$ are needed to identify the best subgroup(s). 
Furthermore, a smaller egonetwork size $n$ requires a larger number of egonetworks under a given power.
In general, under the same scenario, MCB requires more egonetworks than the Wald test because the latter is a single-step test while the MCB will lose some power due to the multiple testing.

Figure \ref{fig:wald_mcb_rep} and Figure \ref{fig:wald_mcb_nrep} show a similar pattern of the sample size with respect to the parameters $\rho_y$ and $n$.
However, the required number of egonetworks in the unique best scenario are larger than that in the multiple best scenario mainly because the distance between the null and alternative is large.

\section{Simulation Study} \label{sec:simul}
To evaluate the finite sample performance of the MCB procedure to identify the best subgroup(s), we conducted a simulation study under different heterogeneity scenarios of the spillover effects, with one or multiple best subgroups and with different differences between the subgroups.
\subsection{Simulation setting}\label{sec:simu_set}

We generate the network data under an ENRT design by the following steps:
1). Generate $K=5000$ egonetworks with $n=5$ social network members within each egonetwork;
2). Randomly assign the $K$ index participants into $H=4$ subgroups with probability $\bm{g} = (g_1,\dots,g_H)=(1/H, \dots, 1/H)$;
3). Randomly assign treatment to all the index participants with probability $p=0.5$;
4). Generate $Y_{ik}$ based on the model \eqref{eq:model}.
The variance of the random effect and residual error are $\sigma_u^2=4$ and $\sigma_e^2=1$ such that $\sigma^2 = 5$ and $\rho_y=0.8$. Let $(\zeta_1,\dots,\zeta_H) = (1,\dots, 1)$. The type I error for computing the simultaneous confidence intervals and test is $\alpha = 0.05$.

We consider the following four scenarios:
\begin{description}
    \item[Scenario 1] Single best subgroup: $\bm{\delta} = (1,2,3,4)$. The subgroup 4 is the best. Hence the differences with the true best of others are $\bm{\Delta}_{\delta} = (\delta_h -\max_{j\neq h} \delta_j, h=1,\dots,H)= (-3, -2, -1, 1)$.
    \item[Scenario 2] Multiple best subgroups: $\bm{\delta} = (1,2,4,4)$. The subgroups 3 and 4 are the best. Hence $\bm{\Delta}_{\delta} = (-3, -2, 0, 0)$.
    \item[Scenario 3] Multiple best subgroups: $\bm{\delta} = (1,3.5,4,4)$. The subgroups 3 and 4 are the best. Hence $\bm{\Delta}_{\delta} = (-3, -0.5, 0, 0)$. Unlike Scenario 2, the difference between the best subgroup(s) and the second-best subgroup(s) is smaller in this scenario. 
\end{description}

\subsection{Results}


\begin{table}
    \begin{tabular}{p{1.2cm} p{0.3cm}p{0.5cm}p{0.8cm}p{2.4cm}p{0.8cm}p{1.3cm}p{0.8cm}p{0.8cm}p{0.8cm}p{0.8cm}p{0.8cm}p{0.8cm}}
        \toprule
        \multirow{2}{*}{Scenario} &\multirow{2}{*}{$\bm{\delta}$} & \multirow{ 2}{*}{True} & \multirow{ 2}{*}{Bias} & \multirow{2}{*}{StdE(eStdE)} & \multirow{2}{*}{$C(\alpha)$} & \multicolumn{6}{c}{MCB}  & {Wald}  \\
        \cmidrule{7-12} 
       & & & & & & Compare & $\text{Pval}^*$ & $\text{Pval}$ & $C(\alpha)^*$ & $\text{$C(\alpha)$}$ & $\text{Power}$ & \text{Power} \\ 
        \midrule
       \multirow{4}{*}{1} & $\delta_1$ & 1 & 0.013 & 0.052 (0.058) & 0.952 & $\delta_1-\delta_4$ & 0.011& \multirow{4}{*}{0.008} & 0.964  &\multirow{4}{*}{0.953} & \multirow{4}{*}{0.844} & \multirow{4}{*}{1.000}\\
       & $\delta_2$ & 2 & 0.038 & 0.052 (0.051) & 0.956 & $\delta_2-\delta_4$ & 0.017 & & 0.971 &  & & \\
       & $\delta_3$ & 3 & 0.050 & 0.052 (0.058) & 0.948 & $\delta_3-\delta_4$ & 0.061 & &0.970 &  & & \\
       & $\delta_4$ & 4 & 0.039 & 0.052 (0.059) & 0.951 & $\delta_4-\delta_3$ & 0.982& &0.986 &  & & \\
    \midrule
        \multirow{4}{*}{2} & $\delta_1$ & 1 & 0.043 & 0.052 (0.055) & 0.946 & $\delta_1-\delta_4$ & 0.015 & \multirow{4}{*}{0.008} & 0.956 &\multirow{4}{*}{0.948} & \multirow{4}{*}{0.890} & \multirow{4}{*}{0.998} \\ 
        & $\delta_2$ & 2 & 0.013 & 0.052 (0.050) & 0.955 & $\delta_2-\delta_4$ & 0.042 & & 0.965 & & & \\
        & $\delta_3$ & 4 & 0.051 & 0.052 (0.057) & 0.950 &  $\delta_3-\delta_4$ & 0.962 &  &0.981 & & & \\
        & $\delta_4$ & 4 & 0.022 & 0.052 (0.051) &0.951 & $\delta_4-\delta_3$ & 0.964 & &0.978 & & & \\
    \midrule
        \multirow{4}{*}{3} & $\delta_1$ & 1 & 0.023 & 0.052 (0.058) & 0.953 & $\delta_1-\delta_4$ & 0.013 &\multirow{4}{*}{0.012}  & 0.962 &\multirow{4}{*}{0.955} & \multirow{4}{*}{0.821} & \multirow{4}{*}{0.972} \\ 
        & $\delta_2$ & 3.5 & 0.011 & 0.052 (0.050) & 0.948 & $\delta_2-\delta_4$ & 0.078 & & 0.961 & & & \\
        & $\delta_3$ & 4 & 0.032 & 0.052 (0.059) & 0.951 &  $\delta_3-\delta_4$ & 0.955 &  &0.982 & & & \\
        & $\delta_4$ & 4 & 0.011 & 0.052 (0.050) & 0.950 & $\delta_4-\delta_3$ & 0.958 &  &0.986 & & & \\
    \midrule
        \multirow{4}{*}{4} & $\delta_1$ & 2 & 0.016 & 0.052 (0.057) & 0.952 & $\delta_1-\delta_4$ & 0.985 & \multirow{4}{*}{0.992}  & 0.994 &\multirow{4}{*}{0.968} & \multirow{4}{*}{0.976} & \multirow{4}{*}{---} \\ 
        & $\delta_2$ & 2 & 0.024 & 0.052 (0.053) & 0.952& $\delta_2-\delta_4$ & 0.992  & & 0.994 & & & \\
        & $\delta_3$ & 2 & 0.038 & 0.052 (0.050) & 0.948& $\delta_3-\delta_4$ & 0.989 & & 0.993 & & & \\
        & $\delta_4$ & 2 & 0.048 & 0.052 (0.051) & 0.951& $\delta_4-\delta_3$ & 0.990 & & 0.994 & & & \\
    \bottomrule
    \end{tabular}
    \caption{Simulation results for Scenario 1, 2, 3 and 4.
 The Wald power for Scenario 4 is not provided, as it is undefined given the null hypothesis is true.}
    \label{tab:simu}
\end{table}


Table \ref{tab:simu} shows the results for the four scenarios. 
$C(\alpha)$ is the coverage rate indicating the proportion of the 95\% confidence intervals estimated by the GEE model that covered the true parameters.  \textit{StdE} is the asymptotic standard deviation calculated from lemma \ref{lem:var_cat} and is averaged across 1000 simulations. We also report its empirical counterpart (\textit{eStdE}). $C(\alpha)^*$ is the coverage rate for a single comparison for the MCB confidence intervals.  \textit{Pval}* is the BH adjusted individual p-value for a single comparison defined in Section \ref{sec:p-value}. 
Both \textit{Pval} and \textit{Pval}* are averaged over 1000 repetitions. 
Although we would prefer to compare the numerical coverage rate to the theoretical value,
by theorem \ref{thm:coverage},  the theoretical value is unavailable for the multiple best scenarios although its lower bound is guaranteed to be $1-\alpha$. The performance of the coverage rates of MCB algins with Theorem \ref{thm:coverage}. The coverage rate converges to 95\% when the difference between the best subgroup(s) and the second-best subgroup(s) is substantial, While for other cases the coverage rate is bounded below by $95\%$

It is seen that the bias of the estimate for $\bm{\delta}$ is very small. When the sample size is large enough (in our example the number of networks is 5000), the empirical variance convergences to the theoretical variance, and hence the $1-\alpha$ confidence intervals cover the true parameter with probability around $95\%$.
But for the MCB test,   the coverage rate of each comparison $\delta_h-\max_{j\neq h} \delta_j$ is larger than the overall coverage rate, because the overall coverage rate is in fact the probability that the simultaneous confidence intervals cover all the true parameters. The power of the Wald test, in scenarios with heterogeneity, 
convergences to the asymptotic power (the theoretical value is 1) when the sample size is large enough.
The MCB power is always lower than that of the Wald test, as proven in the previous section.

The results for Scenario 2 and Scenario 3 in Table \ref{tab:simu} show the Wald test power will decrease if the difference between the null and alternative becomes smaller, which is consistent with asymptotic results. 
The MCB power in Scenario 2 is larger than that in Scenario 1, because there are only two non-best groups in Scenario 2 (this means intuitively it only needs to reject the null twice in stead of three times as in scenario 1), even through the difference between the null and alternative becomes smaller.
In Scenario 4, the p-value of MCB approaches 1 implying non-existence of heterogeneity. Furthermore, the coverage rate is significantly above $95\%$ because there is no difference between any two subgroups, making it the most challenging scenario for identifying the best subgroup(s).

\section{Illustrative Example} \label{sec:data}

HIV prevention intervention studies have shown that injection drug use and sexual transmission among the injection drug users are major sources of HIV infection \citep{davey2011results, tobin2011step}. Here, we analysis the data from the STEP into Action study \citep{tobin2011step}, an ENRT which randomized active injection drug users, referred to as index participants, to receive a training to be health educators and focus outreach specifically to individuals in their social network who inject or are sexual partners, referred here to as social network members (SNMs). 
Spillover of HIV behavioral interventions is common due to peer influence. For instance, educating one drug user about risky behaviors may affect the drug injection and sexual behaviors of their drug-sharing or sex partners.
Additionally, some individuals who receive the intervention, may be more likely to influence others' behaviors. Identifying and targeting these influential individuals would improve the effectness of HIV interventions.
Therefore, we will estimate the spillover effect and identify the key influencers who can greatly affect their network members' behaviors.

The STEP trial was conducted from Mach of 2004 to March of 2006. Two hundred and twenty seven index participants are randomly assigned to the intervention group (114) and control group (113).
Each index participant is asked to invite between one and five SNMs into the study to enable assessment of spillover effects. In total, 424 individuals are recruited as network members.
The outcomes of both index participants and network members are measured by index injection and sex risk behaviors. Follow-up periods were at 6, 12 and 18 month after the baseline session.
In the analysis of this data, the 18 months follow-up period is used.
The demographic characteristics serving as potential covariates to assess heterogeneity  include age, race, education, homelessness and current employment status. 

Here, the outcome is measured by a summary of 
injection and drug splitting risks. 
We consider the age-education as the categorical variable $X_{ik}$,
there are six levels for age-education: young - elementary school (level 1), young - high school (level 2), young - college (level 3), midage - elementary school (level 4), midage - high school (level 5) and midage -college (level 6). We also explored other categorical variables (see the results in supplementary material S4).

Table \ref{tab:real_gee} represents results of the mixed effect regression model estimating the spillover effects ($\delta_h, h=1,\dots, H$) for subgroups defined by age and education levels. 
The findings indicate significant spillover effect in risky behavior for subgroups with age-education levels 1, 4 and 6.
The Wald test for spillover effects with null hypothesis $H_0: \delta_1 = \dots = \delta_H=0$,
confirms the existence of significant spillover effects.

Table \ref{tab:mcb_individual} shows the results of MCB test. Since the 
outcome is the HIV risk score, smaller values indicate better spillover effects.
The Wald test for heterogeneity of spillover effect ($H_0: \delta_1 = \dots = \delta_H$) shows at least one subgroup that significantly differs from the others. 
The simultaneous confidence intervals reveal that subgroups with age-education levels 1 and 2 are significantly inferior to the best groups with age-education levels 3, 4, 5 and 6, whose simultaneous confidence intervals cover 0, while the confidence intervals for other levels are greater than 0.

\begin{table}[ht]
\centering
\begin{threeparttable}
\begin{tabular}{lrrrrc}
  \toprule
coefficient & estimate & std.error & t value & p-value & Wald\\ 
  \midrule
age\_edu1 ($\zeta_1$) & 2.096 & 0.089 & 23.557 & 0.000 & \multirow{6}{*}{---}\\ 
  age\_edu2 ($\zeta_2$) & 2.108 & 0.118  & 17.921 & 0.000 & \\ 
  age\_edu3 ($\zeta_3$) & 1.903 & 0.183  & 10.380 & 0.000 & \\ 
  age\_edu4 ($\zeta_4$) & 2.092 & 0.114 & 18.295 & 0.000 & \\ 
  age\_edu5 ($\zeta_5$) & 1.849 & 0.152  & 12.171 & 0.000 & \\ 
  age\_edu6 ($\zeta_6$) & 2.014 & 0.195  & 10.338 & 0.000 & \\ 
\midrule
  age\_edu\_treat1 ($\delta_1$) & -0.388 & 0.134  & -2.890 & 0.004 &\multirow{6}{*}{0.000}\\ 
  age\_edu\_treat2 ($\delta_2$) & -0.162 & 0.160  & -1.014 & 0.312 \\ 
  age\_edu\_treat3 ($\delta_3$) & -0.276 & 0.269  & -1.025 & 0.307 \\ 
  age\_edu\_treat4 ($\delta_4$) & -0.446 & 0.141  & -3.160 & 0.002 \\ 
  age\_edu\_treat5 ($\delta_5$)& -0.340 & 0.203  & -1.671 & 0.097 \\ 
  age\_edu\_treat6 ($\delta_6$)& -1.008 & 0.300  & -3.361 & 0.001 \\ 
   \bottomrule
\end{tabular}
\end{threeparttable}
\caption{Mixed effect model estimation for subgrous defined by age-education.} 
\label{tab:real_gee}
\end{table}

\begin{table}[ht]
\centering
\begin{threeparttable}
\begin{tabular}{lrrrrrrc}
  \toprule
 contrast  & estimate & std.error & t value & p-value & $L_h$ & $U_h$ & Wald\\ 
  \midrule
 $\delta_1 - \min_{j\neq 1}\delta_j$ & 0.620 & 0.328 & 1.887 & 0.083 & 0.000 & 1.324 & \multirow{6}{*}{0.021} \\ 
$\delta_2-\min_{j\neq 2}\delta_j$  & 0.845 & 0.340 & 2.487 & 0.022 & 0.000 & 1.574\\ 
$\delta_3-\min_{j\neq 3}\delta_j$  & 0.731 & 0.403 & 1.814 & 0.096 & -0.133 & 1.595 \\ 
$\delta_4-\min_{j\neq 4}\delta_j$  & 0.561 & 0.331 & 1.694 & 0.120 & -0.149 & 1.272\\ 
 $\delta_5 - \min_{j\neq 5}\delta_j$  & 0.668 & 0.362 & 1.844 & 0.091 & -0.108 & 1.444 \\ 
$\delta_6-\min_{j\neq 6}\delta_j$  & -0.561 & 0.331 & 1.694 & 0.120 & -1.272 & 0.149\\  
   \bottomrule
\end{tabular}
\end{threeparttable}
\caption{P-value and simultaneous confidence intervals of MCB test for subgroups defined by age-education.}
\label{tab:mcb_individual}
\end{table}


\section{Conclusion and Discussion} \label{sec:discussion}

This paper develops a methodology for assessing the heterogeneity of spillover effects in egocentric network-based randomized trials (ENRTs), with a focus on identifying subgroups of key influencers who maximize the impact of interventions on their social network neighbors. 
By leveraging a regression-based framework and the Multiple Comparison with the Best (MCB) test, we provide a statistical approach to identify the subgroup of index participants who would have the highest average spillover effect on their network members. 

In addition, we generalize the standard MCB \citep{hsu1996multiple} in several ways. First, \citet{hsu1996multiple} assumed an equicorrelation structure for $\bm{R}_h$, and required $\bm{R}_h$ to be identical across subgroups. Here, we extend the MCB method to a general covariance structure. Furthermore, to compute critical values for each subgroup, we use a computationally efficient double integral approach, avoiding the need for numerically simulating a $H$-dimensional distribution for each subgroup.
Second, \citet{hsu1996multiple} claimed exact coverage rate of $1-\alpha$ for MCB simultaneous confidence intervals under the assumption that the difference between the best and the second best goes to infinity (without rigorous proof). Our approach relaxes this assumption, requiring only the presence of the unique best subgroup while maintaining the exact coverage rate under the proposed framework and estimators. 
Lastly, we generalize the concept of power in the standard MCB framework to accommodate scenarios with multiple best subgroups. 


This work is affected by some limitations. Our study design assumes uniform network sizes, which may not reflect real-world variability. Future research could explore methods to accommodate varying network sizes effectively.
Another limitation is the neighborhood interference assumption, which may be violated when the outcome is measured long after the treatment and network members of different networks are connected. Our non-overlapping egonetworks assumption may also be violated when  some network members are connected to multiple index participants. In this case, spillover exposure may not be binary and spillover effects may be defined for different contrasts with potentially different best subgroups for each contrast.  


\section*{Acknowledgements}

This research was partially supported by the award R01MH134715 from National Institute of Mental Health of the National Institutes of Health.\vspace*{-8pt}


\makeatletter
\renewcommand\@biblabel[1]{}
\makeatother
 
\bibliography{ref}






\end{document}